\UseRawInputEncoding
\documentclass[journal=jacsat,manuscript=article]{achemso}
\usepackage{chemformula} 
\usepackage[T1]{fontenc} 

\author{Sisira~Suresh}
\affiliation{Department of Physics, University of Ottawa, Ottawa, ON K1N 6N5, Canada}

\author{Orad~Reshef}
\affiliation{Department of Physics, University of Ottawa, Ottawa, ON K1N 6N5, Canada}
\email{orad@reshef.ca}

\author{M.~Zahirul~Alam}
\affiliation{Department of Physics, University of Ottawa, Ottawa, ON K1N 6N5, Canada}

\author{Jeremy~Upham}
\affiliation{Department of Physics, University of Ottawa, Ottawa, ON K1N 6N5, Canada}

\author{Mohammad~Karimi}
\affiliation{School of Electrical Engineering and Computer Science, University of Ottawa, Ottawa, ON K1N 6N5, Canada}

\author{Robert~W.~Boyd}
\affiliation{Department of Physics, University of Ottawa, Ottawa, ON K1N 6N5, Canada}
\alsoaffiliation {Institute of Optics and Department of Physics and Astronomy, University of Rochester, Rochester, NY 14627, USA}

\title[\textsf{achemso}]
  {Enhanced Nonlinear Optical Responses of Layered Epsilon-Near-Zero Metamaterials at Visible Frequencies}
\abbreviations{IR,NMR,UV}

\begin{document}
\begin{abstract}
Optical materials with vanishing dielectric permittivity, known as epsilon-near-zero (ENZ) materials, have been shown to possess enhanced nonlinear optical responses in their ENZ region. These strong nonlinear optical properties have been firmly established in homogeneous materials; however, it is as of yet unclear whether metamaterials with \emph{effective} optical parameters can exhibit a similar enhancement. Here, we probe an optical ENZ metamaterial composed of a subwavelength periodic stack of alternating Ag and Si$\rm{O_2}$ layers and measure a nonlinear refractive index $n_2$ = (1.2~$\pm~0.1)\times10^{-12}~\rm{m^2/W}$ and nonlinear absorption coefficient $\beta$=  (-1.5~$\pm~0.2)\times10^{-5}~\rm{m/W}$ at its effective zero-permittivity wavelength. The measured $n_2$ is $10^7$ times larger than $n_2$ of fused silica and four times larger than that the $n_2$ of silver. We observe that the nonlinear enhancement in $n_2$ scales as $1/(n_0 \mathrm{Re}{[n_0]})$, where $n_0$ is the linear effective refractive index. As opposed to homogeneous ENZ materials, whose optical properties are dictated by their intrinsic material properties and hence are not widely tunable, the zero-permittivity wavelength of the demonstrated  metamaterials may be chosen to lie anywhere within the visible spectrum by selecting the right thicknesses of the sub-wavelength layers. Consequently, our results offer the promise of a means to design metamaterials with large nonlinearities for applications in nanophotonics at any specified optical wavelength.
  
\end{abstract}
\subsubsection{Keywords}
Epsilon-near-zero, metamaterials, nonlinear optics, multilayer stack, nanophotonics
\section{Introduction}
In recent years, much attention has been given to a class of materials with vanishing dielectric permittivity~\cite{Silveirinha2006,Adams2011,Prain2017}. This class of materials, known as epsilon-near-zero (ENZ) materials, has become a topic of interest because of its intriguing optical properties including tunneling of light through arbitrary bends~\cite{Silveirinha2006}, the ability to tailor radiation patterns~\cite{Alu2007}, and its enhanced nonlinear optical response~\cite{Ciattoni2010, Argyropoulos2012, Capretti2015,Reshef2019}. The ENZ condition can be found in naturally occurring materials near their bulk plasma and phonon resonances. Most noble metals exhibit a zero-permittivity behavior in the UV region, near their respective plasma frequencies~\cite{Johnson1972}. Transition metal nitrides such as titanium nitride~\cite{Wen2006} and zirconium nitride~\cite{Naik2011} display their ENZ regime in the visible spectral region. In the near-infrared (NIR) region, doped semiconducting oxides such as tin-doped indium oxide~\cite{Alam2016a} and aluminium-doped zinc oxide~\cite{Caspania} behave as ENZ materials. An ENZ condition is also found in silicon carbide~\cite{Spitzer1959a}, the perovskite strontium titanate~\cite{Kehr2011}, gallium nitride~\cite{harima1998}, and fused silica ($\rm{SiO_2}$)~\cite{Kischkat2012} in the mid-IR range due to phononic resonances. The zero-permittivity wavelength of a given material is dictated by its intrinsic material properties, and hence cannot be used for applications that require that the ENZ condition occurs at some specified wavelength. To address this concern, ENZ metamaterials have been developed for use in the microwave~\cite{Edwards2008}, IR~\cite{Adams2011, Yang2014} and visible~\cite{Subramania2012, Maas2013} spectral regions. In homogeneous Drude materials, the nonlinear enhancement of $n_2$ and $\beta$ has been thoroughly examined as a function of wavelength in the ENZ region~\cite{Reshef2019}. \textcolor{black}{Although some work has been done exploring the nonlinear response of ENZ metamaterials~\cite{Neira2015, Kaipurath2016,Rashed2020}, its dependence as a function of wavelength has yet to be fully characterized.  Doing this allows us to implicitly infer $n_2$ as a function of $\epsilon$ and, thus, interpret the ENZ condition's real contribution to the optical nonlinearity}. Here, we examine the nonlinear optical response of an ENZ metamaterial that is straightforward to fabricate and for which the ENZ condition can be flexibly set to any targeted wavelength region.\textcolor{black}{Though the nonlinear enhancement in homogeneous ENZ materials has been well established, it is not clear whether such an enhancement occurs in metamaterials when the $\emph{effective}$ permittivity vanishes. In homogeneous materials such as tin-doped indium oxide, the nonlinear enhancement can be explained by a shift in the plasma frequency by intense laser excitation, which changes the permittivity~\cite{Alam2016a}. The refractive index then changes according to $\Delta n= \Delta\epsilon/2 \sqrt{\epsilon}$, which has its maximum value at the zero-permittivity wavelength. It has yet to be established whether $\Delta n$ is maximally changed at the effective zero-permittivity wavelength of a metamaterial}. Our work confirms that a metamaterial indeed does exhibit a nonlinear enhancement in its ENZ region, and therefore, ENZ nonlinear enhancement can be placed at any predefined wavelength. We also develop a simple analytic model to explain these results. 
\begin{figure}[htpb]
\centering
\includegraphics[width=0.85\linewidth]{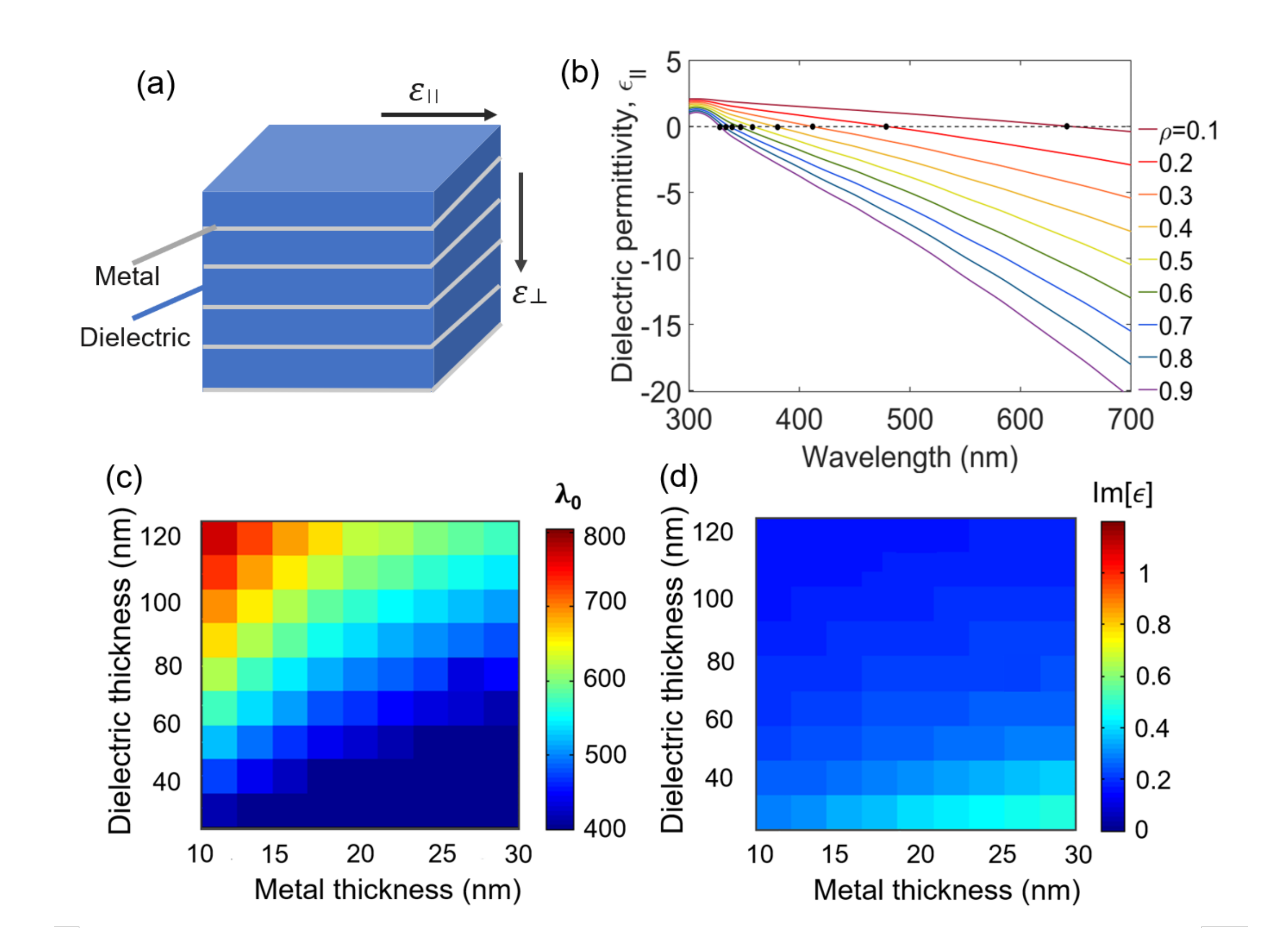}
\caption{(a)~Schematic diagram of a metal-dielectric multilayer stack. (b)~Effective parallel permittivities at normal incidence predicted from EMT for different metallic fill fractions (the black circles denote the zero-crossing wavelength for each fill fraction). (c)~The zero-permittivity wavelength and (d)~the loss of a 5-bilayer Ag-Si$\rm{O_2}$ multilayer stack calculated using TMM as a function of the thicknesses of the Ag and Si$\rm{O_2}$ layers. Note that the zero-permittivity wavelength can be placed anywhere in the visible region.}
\label{fig:Figure1}
\end{figure} 
\par
Our metamaterial is composed of alternating sub-wavelength-thick layers of metal and dielectric materials. A schematic diagram of the metamaterial geometry is shown in Fig.~\textcolor{blue}{\ref{fig:Figure1}(a)}. These metamaterials are capable of exhibiting a zero-permittivity wavelength anywhere within the entire visible spectrum by adjusting the respective thicknesses of the constituent materials\cite{Subramania2012,Shalaev,Omar2012,Newman2015}. 
Provided that the inhomogeneity scale of the composite medium is of sub-wavelength dimensions, effective medium theory (EMT) predicts that the wavelength, $\lambda_0$, at which the permittivity crosses zero can be evaluated from the fill fraction of the constituents in the composite~\cite{Shalaev,Sihvola1999}. Thus, in the limit of sub-wavelength layer thickness, the metal-dielectric multilayer stack can be considered as an effective medium with an effective permittivity for an electric field polarized in the plane of the layers given by $\epsilon_{\rm{\parallel}}= \rho\epsilon_m +(1-\rho)\epsilon_d$, where $\rho$ is the metallic fill fraction and $\epsilon_m$ and $\epsilon_d$ are the permittivities of the metal and the dielectric material, respectively~\cite{Jetp1956}. We selected Ag as the metal because of its small damping constant compared to other noble metals~\cite{West2010}, and $\rm{SiO_2}$ as the dielectric because of its transparency in the visible spectral region~\cite{Palik1998}. In Fig.~\textcolor{blue}{\ref{fig:Figure1}(b)}, the fill fraction is varied from $\rho$ = 0.1 to 0.9, and we observe a blue shift in the zero-permittivity wavelength as the metallic fill fraction increases. Thus, the dependence of the zero-permittivity wavelength on the metallic fill fraction should enable an ENZ metamaterial design that can be situated anywhere in the entire visible spectrum.
\par
\begin{figure}[h!]
\centering
\includegraphics[width=0.85\linewidth]{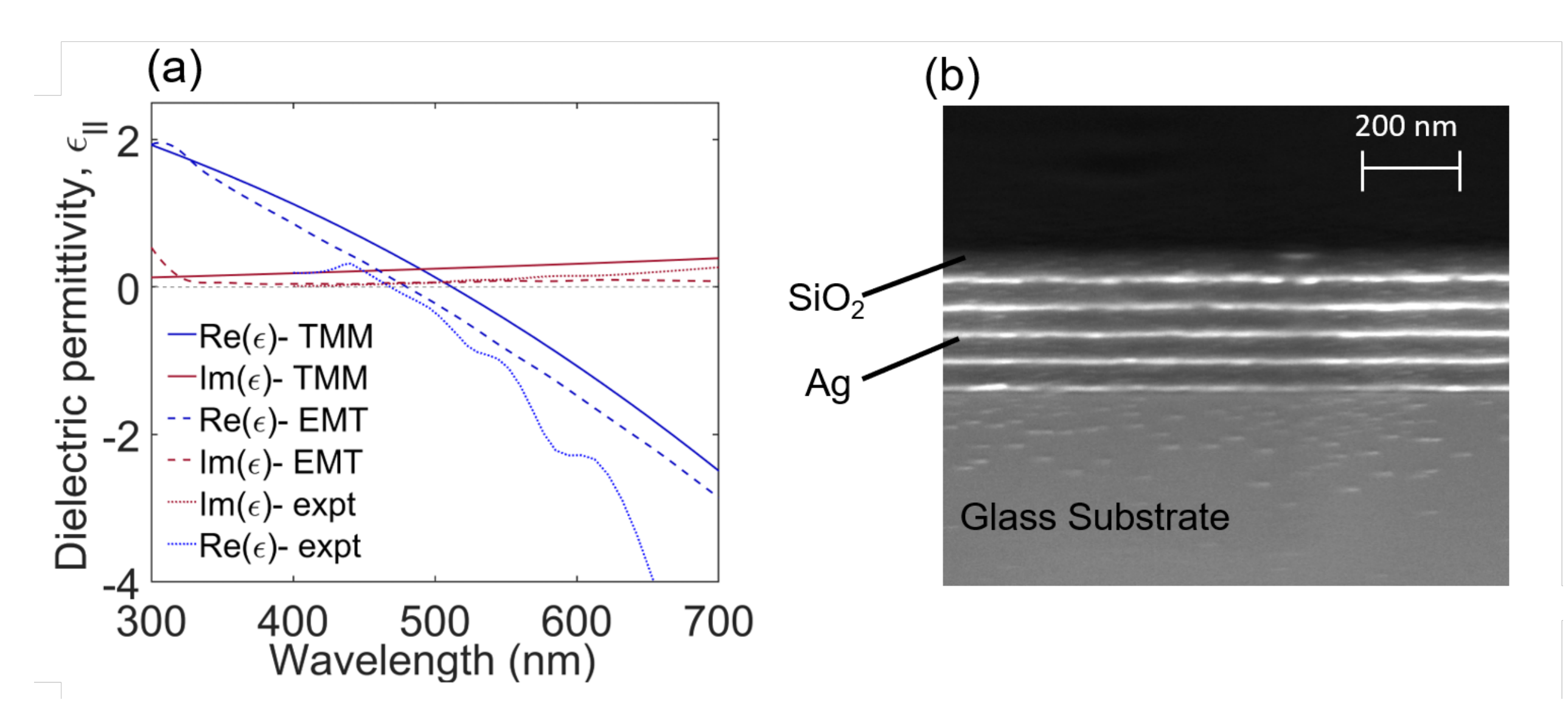}
\caption{(a)~Effective parallel permittivity, $\epsilon_{\parallel}$, at normal incidence calculated using the TMM, EMT, and the measured transmittance for an Ag-$\rm{SiO_2}$ multilayer stack with five bilayers of Ag (16~nm) and $\rm{SiO_2}$ (65~nm). (b)~Cross-sectional image of the fabricated Ag-$\rm{SiO_2}$ multilayer stack taken with a scanning electron microscope.}
\label{fig:Figure2}
\end{figure}

 
Although effective medium theory can reliably predict the ENZ wavelength under many situations, this method is rigorously valid only under limiting conditions, such as vanishingly small layer thickness and an infinitely thick overall medium~\cite{Papadakis2015}. In order to validate our EMT approach, we perform parameter retrieval using the transfer matrix method (TMM) to aid in our design~\cite{Smith2005a}. Using this method, one can solve for the effective refractive index and consequently the complex effective permittivity of a medium. The TMM simulations reveal optimal designs in terms of zero-permittivity wavelength and optical losses (Fig.~\textcolor{blue}{\ref{fig:Figure1}(c)}). We select a design for the Ag-Si$\rm{O_2}$ multilayer stack that has both a desired zero-permittivity wavelength and a small amount of loss, consisting of five bilayers of Ag and $\rm{SiO_2}$ with thicknesses of 16~nm and 65~nm, respectively, for a total thickness of 405~nm. We choose five bilayers because it has been shown that using more than five bilayers produces no appreciable improvement in the nonlinear optical response \cite{Lepeshkin2004}. Figure~\textcolor{blue}{\ref{fig:Figure2}(a)} depicts the dielectric permittivity at normal incidence as a function of wavelength calculated using both the TMM and EMT methods. The optical losses are due to the resistive losses of silver. This geometry corresponds to a metamaterial with an effective zero-permittivity wavelength of 509~nm, with an imaginary part of dielectric permittivity Im$[\epsilon]$ of 0.2.
\par
Having established a preferred design, we fabricated a device for characterization. The Ag and Si$\rm{O_2}$ layers were deposited using electron-beam evaporation on a glass substrate. The deposition rates of Ag and Si$\rm{O_2}$ layers were kept at a low value of 0.1~nm/s in order to maintain film uniformity. To prevent oxidation, the top layer is the Si$\rm{O_2}$ layer. A cross-section of the fabricated sample is shown in Fig.~\textcolor{blue}{\ref{fig:Figure2}(b)}. Our fabricated sample agrees with our design within the usual fabrication tolerances.
\section{Methods}
The linear transmittance of the sample was probed using a collimated supercontinuum source covering the visible to NIR spectral range. We compared the measured transmission spectra to those predicted by TMM simulations for various metal and dielectric layer thicknesses. We found the best agreement with the experimental data for a metal-dielectric multilayer stack with thicknesses of 16~nm for Ag and 56~nm for Si$\rm{O_2}$ (see supplemental materials for more details). The resulting zero-permittivity wavelength occurs at 470~nm (Fig.~\textcolor{blue}{\ref{fig:Figure2}(a)}), which is reasonably close to the predicted zero-permittivity wavelength of our device design (509~nm). \textcolor{black}{The small discrepancy between the target zero-crossing wavelength and that determined from these linear characterization measurements could be attributable to fabrication uncertainties, such as layer composition or variations in thickness, or measurement uncertainties in the linear characterization of the device}.
 \par
\begin{figure}[h!]
\centering
\includegraphics[width=0.85\linewidth]{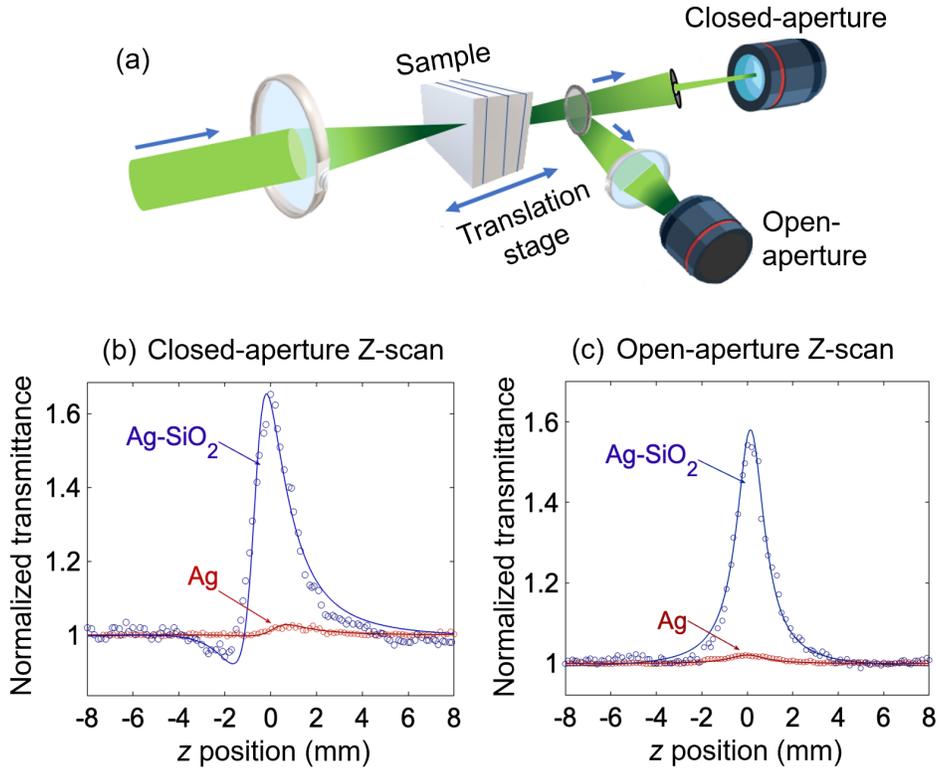}
\caption{(a)~Experimental setup. The Z-scan measurements were performed using 28-ps pulses with a repetition rate of 50 Hz from an optical parametric generator. A spatially filtered Gaussian beam is focused at normal incidence onto the sample by a lens. (b)~The closed- and (c)~open-aperture Z-scan signals at $\lambda$=500~nm for an Ag-SiO$_2$ multilayer stack (blue) and a thin film Ag layer (red) at normal incidence are shown. The solid lines represent theoretical fits to the experimental data.}
\label{fig:Figure3}
\end{figure}

 We characterized the nonlinear optical properties of our sample using the Z-scan technique~\cite{Sheik-Bahae1990}.
  A schematic diagram of the experimental setup is shown in Fig.~\textcolor{blue}{\ref{fig:Figure3}(a)}. 
We used pump pulses with a repetition rate of 50~Hz and a pulse duration of 28-ps from an optical parametric generator. Both closed- and open-aperture measurements were performed for wavelengths ranging from 410~nm to 560~nm. Note that the entire spectral range is in the ENZ region. All the measurements were conducted at normal incidence. \textcolor{black}{As such, we do not expect to excite any surface plasmon polaritons}. Figures~\textcolor{blue}{\ref{fig:Figure3}(b)} and \textcolor{blue}{(c)} show, respectively, representative closed-aperture and open-aperture signals from the Ag-Si$\rm{O_2}$ multilayer stack at $\lambda=500$~nm. The asymmetry in the closed-aperture signal with respect to the focus is due to the significant nonlinear absorption in the sample~\cite{Liu2001,Tsigaridas2009}. We first extracted the imaginary part of the nonlinear phase shift from the open-aperture signal and used this value to calculate the real part of the phase shift from the closed-aperture signal. The extracted values of the real and imaginary nonlinear phase shifts were used in the standard expressions to calculate $n_2$ and $\beta$ (see supplemental materials for details)~\cite{Sheik-Bahae1990}. For comparison, Figs.~\textcolor{blue}{\ref{fig:Figure3}(b)} and \textcolor{blue}{(c)} also show similar measurements performed under the same conditions for a single 16-nm-thick Ag layer. Near the zero-permittivity wavelength, the accumulated nonlinear phase of the multilayer stack is 22 times larger than that of the 16-nm-thick silver layer, even though the multilayer stack contains only 5 times as much silver.

\begin{figure}[!ht]
\centering
 \includegraphics[width=0.85\linewidth]{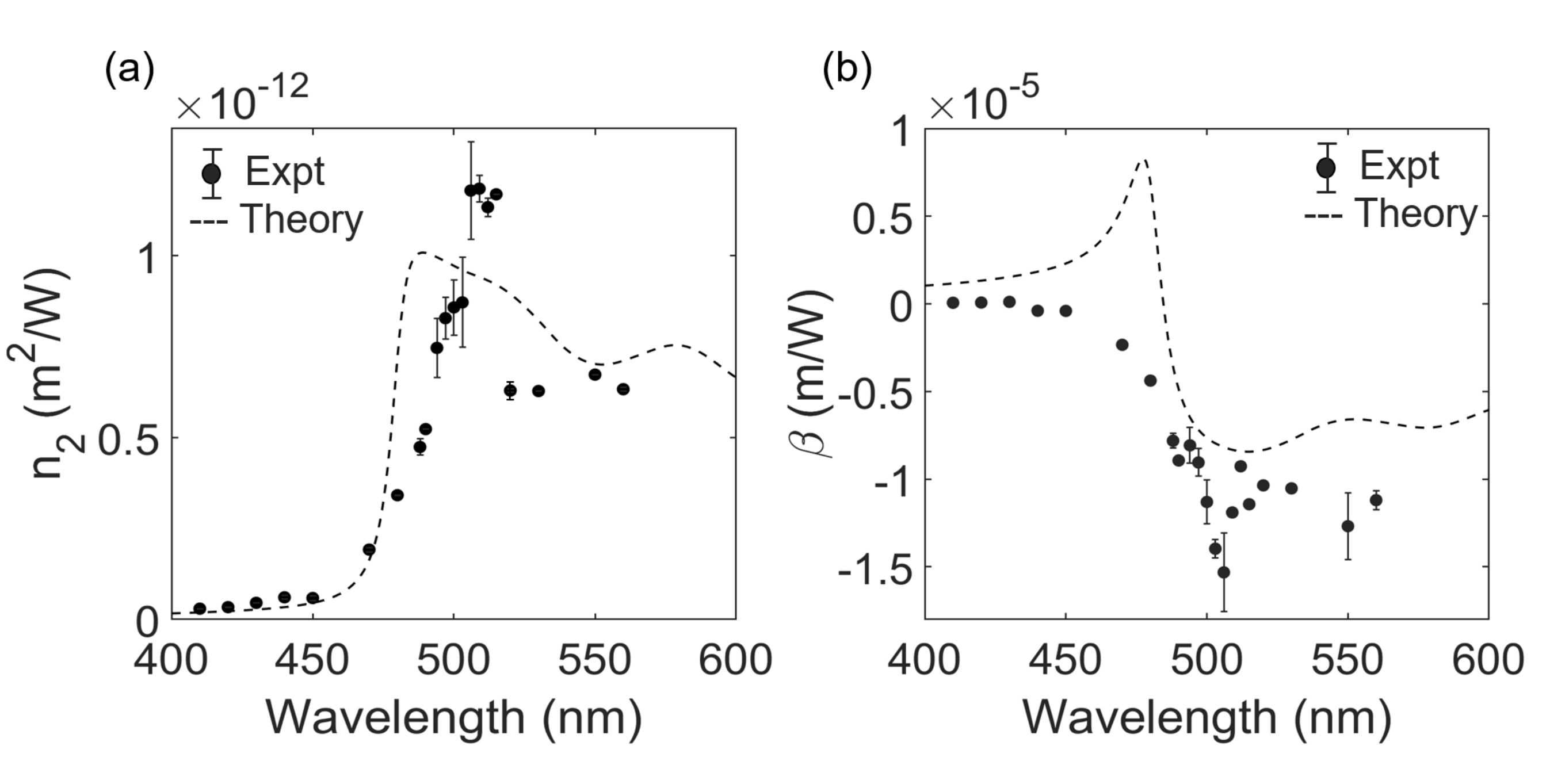}
\caption{(a)~The nonlinear refractive index $n_2$ and (b)~the nonlinear absorption coefficient $\beta$ of the Ag-Si$\rm{O_2}$ multilayer stack as a function of wavelength. The dashed lines correspond to predictions from Eqs.~\eqref{EQ:n2} and \eqref{EQ:beta} without any fit parameters.}
\label{fig:Figure4}
\end{figure}
\par
The nonlinear refractive index $n_2$ and the nonlinear absorption coefficient $\beta$ of the Ag-Si$\rm{O_2}$ multilayer stack are shown as functions of wavelength in Figs.~\textcolor{blue}{\ref{fig:Figure4}(a)} and \textcolor{blue}{(b)}. It is clear that the nonlinear response is enhanced in the ENZ region of the spectrum, peaking at the zero-permittivity wavelength targeted by this metamaterial design.  The maximum measured phase shift at the zero-permittivity wavelength is $0.62\pi\pm0.05$~rad. For the Ag-Si$\rm{O_2}$ multilayer stack, the values of $n_2$ and $\beta$ are (1.2~$\pm~0.1)\times10^{-12}~\rm{m^2/W}$ and (-1.5~$\pm~0.2)\times10^{-5}~\rm{m/W}$, respectively. The peak value of the measured $n_2$ of the Ag-Si$\rm{O_2}$ multilayer stack is $10^7$ times larger than that of fused silica (3.2$\times10^{-20}~\rm{m^2/W}$)~\cite{Boyd2013}, and is four times larger than that of an individual 16-nm-thick silver film (3~$\times10^{-13}~\rm{m^2/W}$). \textcolor{black}{Due to the non-instantaneous nature of the nonlinearity of metals, we would expect to obtain different values for the nonlinear response for different experimental conditions. For example, we expect that performing the same measurements as reported above with shorter pulses would lead to smaller magnitudes of nonlinearity \cite{Boyd2014}. However, by performing our measurement with a narrow-band pulse, we are able to measure the nonlinear response across a broad spectral range spanning over the ENZ wavelength for this sample, confirming the existence of clear nonlinear enhancement due to the zero-permittivity wavelength.}

\section{Results and Discussion}
We model the nonlinearity of the metamaterial stack using the nonlinear EMT~\cite{Boyd1994}. Here, the effective nonlinear susceptibility of the metamaterial stack is the weighted average of the constituent materials. Since $\chi^{(3)}_{\mathrm{SiO_2}}$ is much smaller than $\chi^{(3)}_{\mathrm{Ag}}$, according to EMT, the dominant contribution to $\chi^{(3)}_\mathrm{eff}$ of the metamaterial is from the Ag layers only (\emph{e.g.,} $\chi^{(3)}_\mathrm{eff}\approx\chi^{(3)}_{\mathrm{Ag}}\times\rho$). We assume that $\chi^{(3)}_{\mathrm{Ag}}$ is dispersionless over this spectral range. We measured our single silver layer sample at $\lambda=500$~nm and obtained $\chi^{(3)} = (2.42 + 5.15i)\times10^{-16}~\rm{m^2/V^2}$, in good agreement with previously measured values~\cite{Ma2007a,Yang2003}. The complex nonlinear response $\tilde{n}_2$ of the composite material is given by~\cite{Boyd2013,Sutherland1996,DelCoso2004}
\begin{equation}
\tilde{n}_2 = \frac{3}{4\epsilon_0 c  n_0 \mathrm{Re}[n_0]} \chi_{\rm{eff}}^{(3)}, \label{EQ:Standard_n2}
\end{equation}
where $\epsilon_0$ is the vacuum permittivity and $c$ is the speed of light in vacuum. Eq.~\eqref{EQ:Standard_n2} is related to the nonlinear refraction $n_2$ and the absorption coefficient $\beta$ by the relations
\begin{align}
    n_2 &= \mathrm{Re}[\tilde{n}_2] \label{EQ:n2}
\\
    \beta &= \frac{4\pi}{\lambda}\mathrm{Im}[\tilde{n}_2].
\label{EQ:beta}
\end{align}
We plot these equations in Fig.~\textcolor{blue}{\ref{fig:Figure4}} using the refractive index of our design as calculated by the EMT.

The model shows a strong, wavelength-dependent enhancement at the zero-permittivity wavelength that qualitatively resembles the experimental results. It correctly predicts the location and the maximum nonlinear response to within a factor of two without the need for any fit parameters or additional factors (\emph{e.g.,} the slow-light factor $S=n_g/n_0$, where $n_g$ is the group index~\cite{Monat2010,boydSlowlight}). The discrepancies in the breadth and the magnitude of this enhancement at the peak could likely be attributed to dimension variations between the design and the fabricated device, surface effects, imperfections in the constituent layers introduced during deposition, or our assumption that $\chi^{(3)}_{\mathrm{Ag}}$ is dispersionless in our theoretical model. We note that our model predicts an additional peak for $\beta$ at $\lambda=475$~nm that we do not reproduce in the measurement and currently cannot account for. The qualitative agreement between such a simple theory and the experimental results suggests that this model may be used to predict and design the nonlinear optical response of other ENZ metamaterials.
\par
In order to study the nature of the enhancement of the $n_2$ nonlinear response, we compare the response of the Ag-Si$\rm{O_2}$ multilayer stack directly with that of a single thin film of silver. Given that $\chi^{(3)}_{\rm{eff}}$ $\approx \rho\times \chi^{(3)}_{\rm{Ag}}$, with $\rho<1$, any metamaterial stack composed of Si$\rm{O_2}$ and Ag layers will exhibit a smaller $\chi^{(3)}$ value than that of silver. However, we found that at its peak the magnitude of $n_2$ of the metamaterial is four times that of silver. This observation implies that the ENZ condition increases $n_2$ to exceed the value of silver, despite the silver being ``diluted'' by a material with a lower nonlinearity ($i.e$., Si$\rm{O_2}$). This observation is further validated when comparing $n_2$ and $\beta$ of the ENZ metamaterial at its zero-permittivity wavelength ($\lambda$ = 506~nm) to these same values when $\epsilon_{\rm{eff}} \approx 1$ ($\lambda$ = 410~nm; see Fig.~\textcolor{blue}{\ref{fig:Figure2}(a)}). Here, the magnitudes of $n_2$ and $\beta$ are increased in the ENZ region by factors of 40 and 250, respectively. In addition to this ENZ enhancement, at the zero-permittivity wavelength, the metamaterial has a smaller linear loss than silver (Im$[n_0]=0.3$ vs 3.1, respectively). Consequently, its effective propagation length can be much longer than that of silver (60~nm vs 9~nm), allowing for a much larger accumulation of nonlinear phase~\cite{Bennink1999,Lepeshkin2004}. As shown by the peak-to-valley differences in Fig.\textcolor{blue}{~\ref{fig:Figure3}(b)}, in propagating through a 5-bilayer Ag-Si$\rm{O_2}$ multilayer stack, the beam acquires a nonlinear phase shift that is approximately 22 times larger than that of the individual silver layer (1.53~rad vs 0.068~rad). Therefore, the benefit of using an ENZ metamaterial over a bulk metallic thin film is twofold: due to ENZ enhancement, and due to lowered loss~\cite{Neira2015,Rashed2020,Bennink1999,Lepeshkin2004}.
\par
In conclusion, we have examined the nonlinear optical properties of an ENZ metamaterial realized through use of a metal-dielectric multilayer stack. This work further confirms that the enhancement of the nonlinear optical response that had previously been observed in homogeneous materials at the zero-permittivity wavelength~\cite{Alam2016a,Caspania} occurs also in metamaterials at the zero of the \emph{effective} permittivity~\cite{Neira2015, Kaipurath2016,Rashed2020}. We have observed that these materials produce a large nonlinear optical response and that the dominant mechanism for enhancing this response is the factor $1/(n_0 \mathrm{Re}[n_0])$. The ability to obtain strong nonlinearities at designated optical frequencies makes these metamaterials a flexible platform for applications in nonlinear optics.
\par
There exists a broad variety of nonlinear optical phenomena, of which only the Kerr effect and saturable absorption were directly examined in this work. The investigation of other such nonlinear responses~\cite{Luk2015,Yang2019} and their potential enhancement in ENZ metamaterials certainly warrants further study. The fact that this metamaterial geometry is inherently anisotropic could be seen as  an advantage for certain future applications and be the topic of future study.

\section{Supporting Information}
Supporting information:
This material is available free of charge via the internet at http://pubs.acs.org. Linear characterization methods, beam cleaning procedures, retrieval of nonlinear optical coefficients, peculiar features of open- and closed-aperture z-scan signals, nonlinear phase shift, and additional experimental details are provided. 

\begin{acknowledgement}
This work was supported in part by the Canada First Research Excellence Fund, the Canada Research Chairs Program, and the Natural Sciences and Engineering Research Council of Canada (NSERC [funding reference number RGPIN/2017-06880]). R.W.B. acknowledges support from DARPA (grant No. W911NF- 18-0369) and ARO (Grant W911NF-18-1-0337). O.R. acknowledges the support of the Banting Postdoctoral Fellowship from NSERC. Fabrication in this work was performed at the Centre for Research in Photonics at the University of Ottawa (CRPuO). 
\end{acknowledgement}


\pagebreak

\setcounter{page}{1}
\renewcommand\thesection{S\arabic{section}} 
\setcounter{section}{0}
\renewcommand\thefigure{S\arabic{figure}}   
\setcounter{figure}{0}  

{\Large \bf \noindent Supplementary Information: \\ Enhanced Nonlinear Optical Responses of Layered Epsilon-Near-Zero Metamaterials at Visible Frequencies}

\par
This document provides supporting information to ``Enhanced nonlinear optical responses of layered epsilon-near-zero metamaterials at visible frequencies''. In Sec.1, we show the linear characterization methods of the sample. In Sec.2, we describe the experimental procedure to ensure that the intensity distribution of the beam is Gaussian for the Z-scan measurements. In Sec.3, we explain the equations used for the retrieval of nonlinear coefficients $n_2$ and $\beta$ from Z-scan measurements. In Sec.4, we illustrates a change in sign of the nonlinear absorption in the operational wavelength range in the open-aperture Z-scan signals and in Sec.5, we demonstrates asymmetry in the closed-aperture Z-scan signals in the operating wavelength range. Finally in Sec.6, we present the nonlinear phase shift as a function of irradiance.

 \section{1. Linear characterization of the sample}\label{Section_A}
 \par
\begin{figure}[h!]
\centering
\includegraphics[width=12cm, height=6cm]{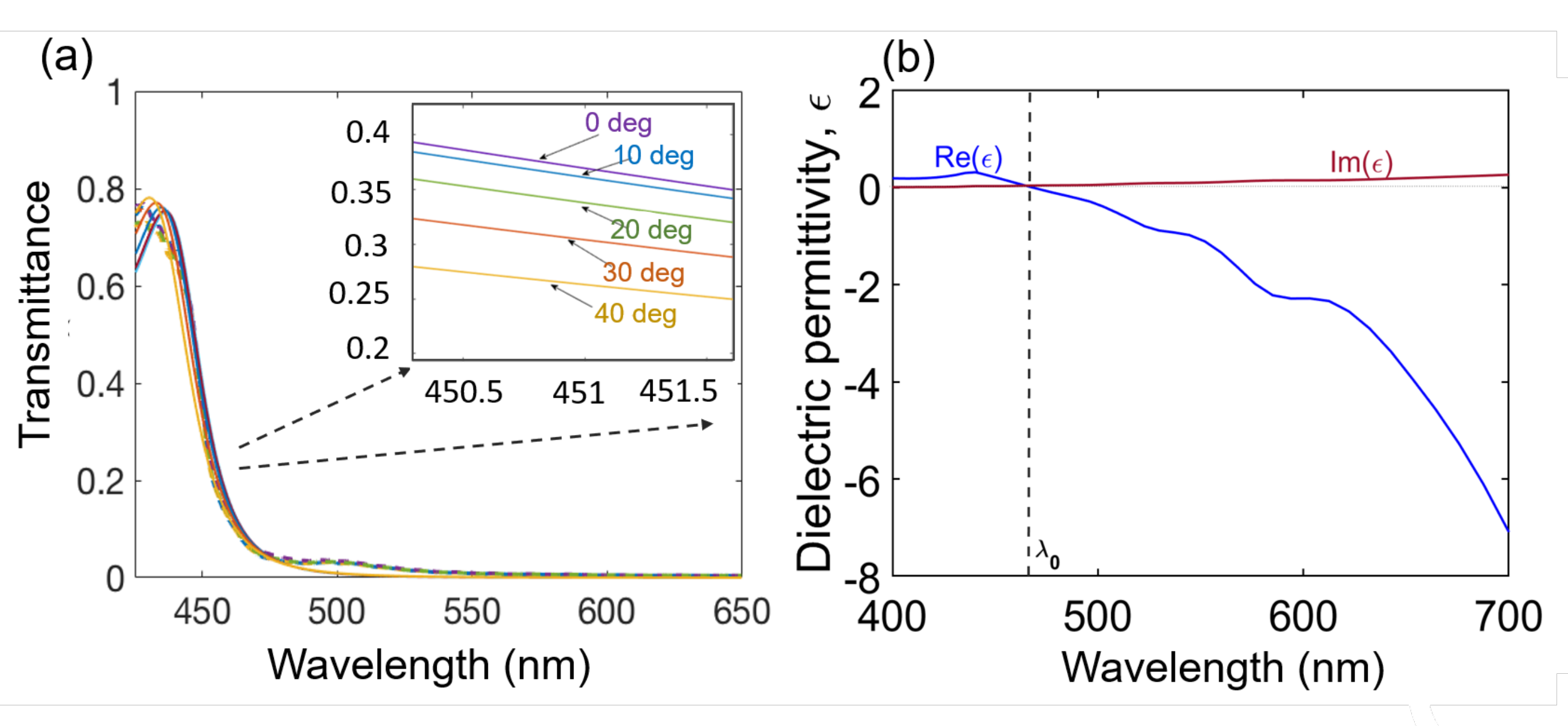}
\caption{(a)~Transmission spectra of the Ag-SiO$_2$ multilayer stack for p-polarization. The inset shows a zoomed-in view of the simulation spectra around 450~nm. The solid lines are the TMM simulation data and the dashed lines are the measurements. Colours represent different angles of incidence. (b)~Dielectric permittivity of the Ag-SiO$_2$ multilayer stack calculated from transmission measurements.}
\label{fig:Sup_Figure1}
\end{figure}

 
The linear response of the Ag-Si$\rm{O_2}$ multilayer stack is characterized by a transmission measurement at varying degrees of incidence. We compared the measured transmission spectra to those expected from the TMM simulations for several thicknesses of metal and dielectric layers in the stack. We found that the TMM simulations are consistent with the experimental data for a metal-dielectric multilayer stack with thickness of 16~nm of silver and 56~nm of Si$\rm{O_2}$, as shown in Fig.~\ref{fig:Sup_Figure1}

\section{2. Beam cleaning}\label{Section_B}
\par
The Z-scan theory is based on the assumption that the incident beam is Gaussian (TEM00). Sometimes, the diffraction effects in the laser resonator cavity induces a change in the beam mode as it propagates outside the laser resulting in a non-Gaussian beam profile. The laser source used in our experiment produces a non-Gaussian beam. We adopted a spatial filtering technique to remove random fluctuations and higher order modes from the beam profile. The spatial filter set-up consists of a lens and  a pinhole aperture attached to an $x-y-z$ positioning mechanism. By removing the higher order spatial modes inside the laser pulse, the spatial filter allows only Gaussian mode (TEM00) to pass through. Figure~\textcolor{blue}{~\ref{fig:Sup_Figure1}(a)} and \textcolor{blue}{(b)} shows the laser beam profile before and after the spatial filter, respectively. 
\par
\begin{figure}[h!]
\centering
\includegraphics[width=12.5cm, height=8.5cm]{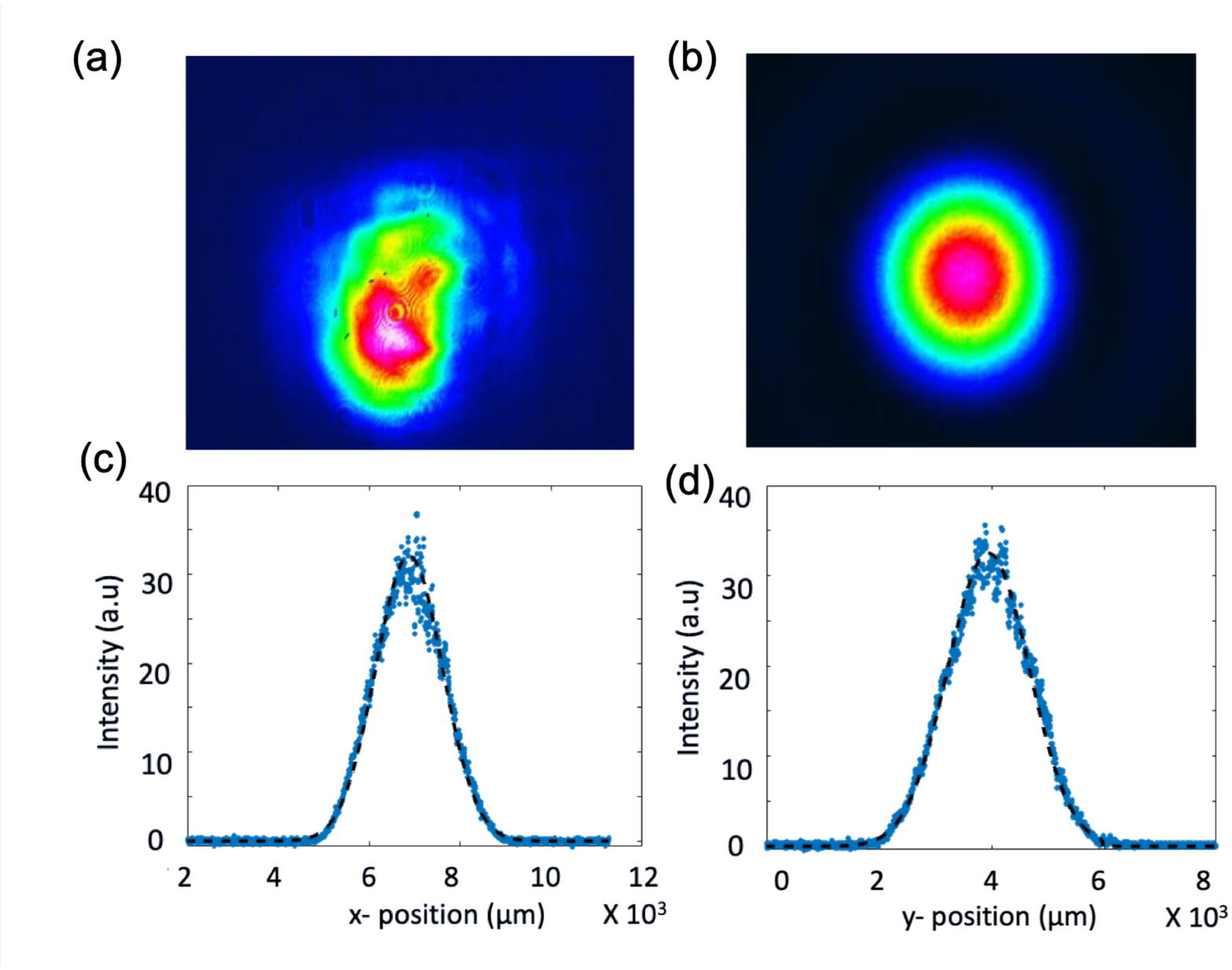}
\caption{(a)~Gaussian beam intensity profile (a) before and (b) after the spatial filter. Gaussian fit of the experimental beam profile in the (c)~$x$- and (d)~$y$- directions.}
\label{fig:Sup_Figure2}
\end{figure}

 
Beam ellipticity $e$ is another parameter which is of great importance in the Z-scan measurements. Large ellipticity values of the beam can badly affect the signal shape in Z-scan measurements. Wicksted $et~al.$~\cite{Mian1996} studied the effects of beam ellipticity on Z-scan measurements and noticed that with an increase in ellipticity, both the peak and valley reduce from their symmetric maxima and minima that were attained in the circular limit ($e=1$). We measure the beam ellipticity at different positions along the propagation length after spatial filtering the beam. Our filtered beam presents a good fit to a Gaussian as shown in Fig.~\textcolor{blue}{\ref{fig:Sup_Figure2}(c)} and \textcolor{blue}{(d)} in $x$ and $y$ directions, respectively.

In the $x$-direction, the Gaussian fit to the experimental beam profile is 90.15\% and in the $y$-direction it is 91.98\%. The $1/e^2$ diameter of the input beam is 2959~$\mu m$ in the $x$-direction and 2909~$\mu m$ in the $y$-direction. From these values, we measure the ellipticity of the beam as 1.01.

\section{3. Retrieval of nonlinear coefficients $n_2$ and $\beta$ from Z-scan measurements}\label{Section_B.2}
\par 
The nonlinear absorption coefficient $\beta$ can be determined directly from the open-aperture measurements. In the open-aperture measurements, the aperture is removed and the total transmittance is collected using a lens. The variations in the transmission are largest at the focus and smallest away from it. For a beam with a Gaussian spatial profile, the transmission in the case of open-aperture is given by 
\begin{equation} \label{eq:OAeq}
{T_{\textrm{OA}}= 1- \frac{\Delta \Psi} {1+ (z/z_0)^2}},
\end{equation}
\newline
where $z_0$ is the Rayleigh range and $\Delta\Psi$ is the imaginary part of nonlinear phase shift given by \cite{Sheik-Bahae1990}
\begin{equation} \label{eq:betaeq}
\Delta\Psi=\frac{\beta I_0 L_{\rm{eff}}}{2 \sqrt{2}},
\end{equation}
where $\beta$ is the nonlinear absorption coefficient, $I_0$ is the peak intensity and $L_{\rm{eff}}$ is the effective sample thickness.
\par 
In the closed-aperture measurement, only a part of the intensity transmitted through the sample is collected. Thus, variations in the transmission correspond to both nonlinear absorption and nonlinear refraction. In the case of closed-aperture, the transmission after the aperture is given by \cite{Sheik-Bahae1990}
\begin{equation}  
T_\textrm{CA}= 1+ \frac{4(z/z_0)\Delta \Phi} {(1+ (z/z_0)^2)(9+ (z/z_0)^2)} -\frac{2((z/z_0)^2 +3)\Delta\Psi} {(1+ (z/z_0)^2)(9+ (z/z_0)^2)}
\label{eq:CAeq}
\end{equation}
\medskip
where $\Delta\Phi$ is the real part of nonlinear phase shift. $\Delta\Phi$ is related to the nonlinear refractive index $n_2$ by
\begin{equation} \label{eq:n2eq}
\Delta\Phi= (1-S)^{0.25} k n_2 I_0 L_{\rm{eff}},
\end{equation}
where $k=2\pi/\lambda$ and $S$ is the transmission through the aperture.
\newline
Therefore, by fitting the experimentally obtained open-and closed-aperture Z-scan transmission profiles with Eqs.~\eqref{eq:OAeq} and \eqref{eq:CAeq}, the nonlinear parameters $n_2$ and $\beta$ can be found.

\section{4. Change in sign of the nonlinear absorption}\label{Section_C}
\par
\begin{figure*}[h!]
\centering
\includegraphics[width=13cm, height=6cm]{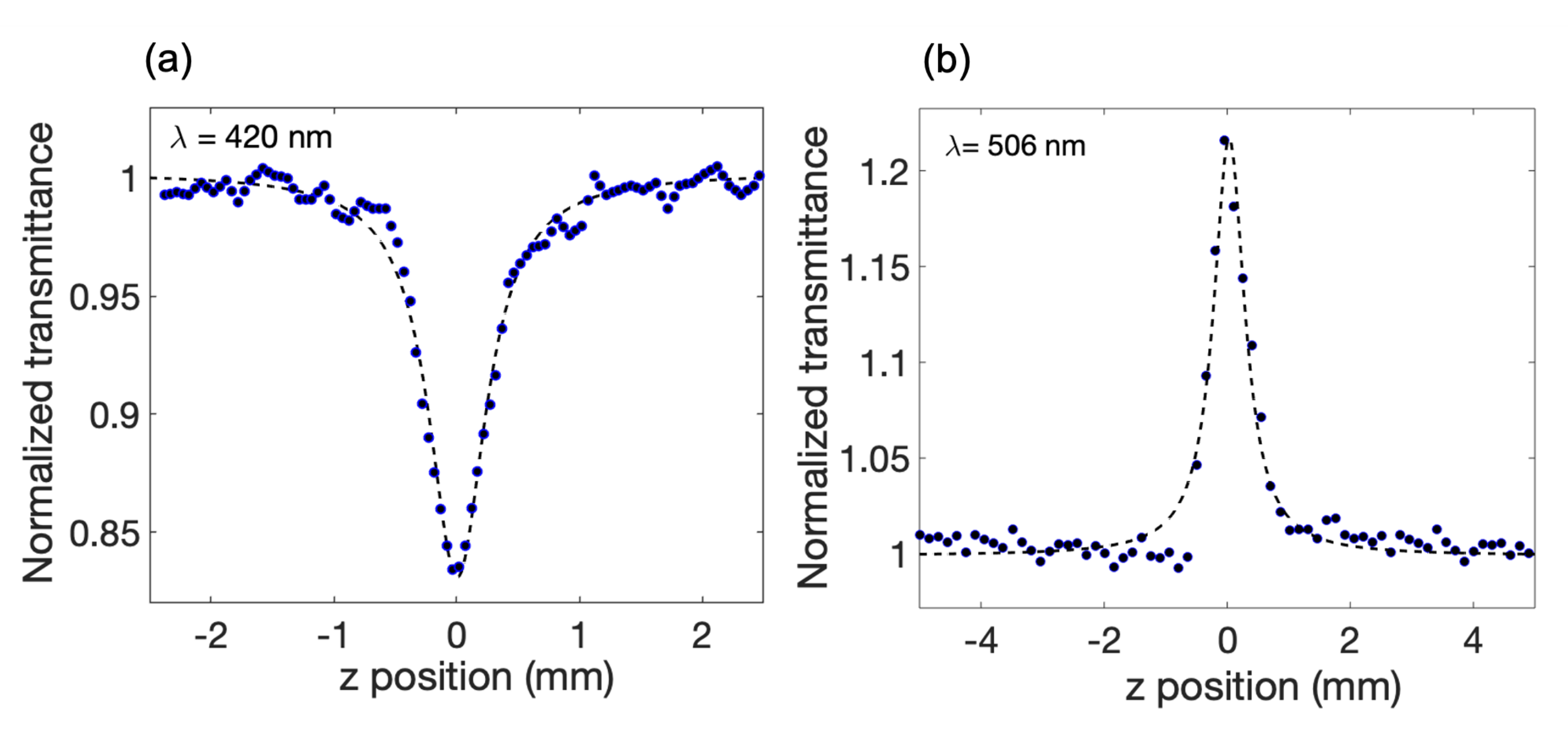}
\caption{Open-aperture Z-scan signal showing a transition from reverse saturable to saturable absorption. (a)~Reverse saturable absorption at 420~nm corresponding to a $\beta=7.4~\times10^{-8}~\rm{m/W}$ (b)~Saturable absorption at 506~nm corresponding to a $\beta=-1.5\times10^{-5}~\rm{m/W}$. We observed reverse saturable absorption for all operating wavelengths below 440~nm and saturable absorption for all wavelengths above 440~nm. The intensity at the focus was $300~\rm{MW/cm^2}$.}
\label{fig:supp_fig3}
\end{figure*}

 
Another interesting result is the change in sign of the nonlinear absorption coefficient as a function of wavelength. The change from reverse saturable absorption (RSA) to saturable absorption (SA) is observed at 440~nm. This is shown in Fig.~\ref{fig:supp_fig3}. Below 450~nm RSA is observed and above 450~nm SA is observed. The physical mechanism behind this sign change is of interest for future research.
\section{5. Asymmetry in the closed-aperture Z-scan signal}\label{Section_D}

\begin{figure}[h!]
\centering
\includegraphics[width=18cm,height=6.3cm]{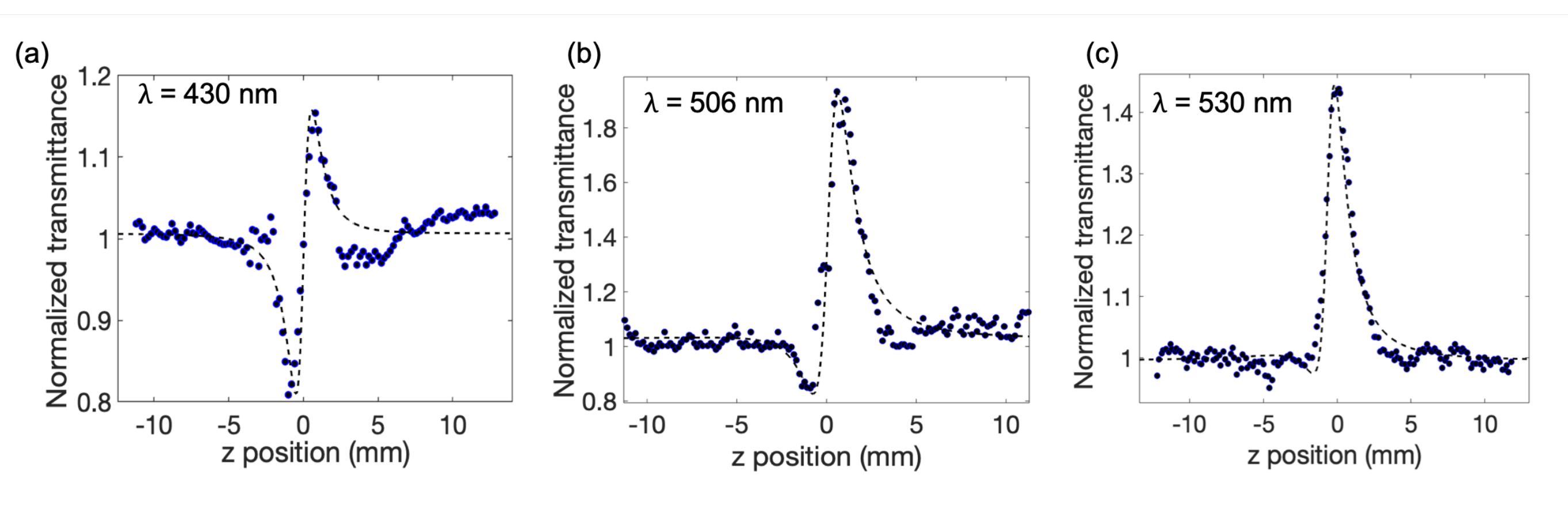}
\caption{Closed-aperture Z-scan signal (a)~at 430~nm (away from zero-permittivity wavelength, towards the blue side of spectrum) (b)~at 506~nm (at the zero-permittivity wavelength) (c)~at 530~nm (away from zero-permittivity wavelength, towards the red side of the spectrum).}
\label{fig:Supp_fig4}
\end{figure}

Figure~\ref{fig:Supp_fig4} shows the closed-aperture Z-scan signal measured at different wavelengths at and away from the zero-permittivity wavelength. As we can see from Fig.~\textcolor{blue}{\ref{fig:Supp_fig4}(a)}, at a shorter wavelength the closed-aperture Z-scan signal is symmetric with respect to the focus ($\textit{i.e}$, at z=0). As we move towards the longer wavelength region, the closed-aperture signal loses its symmetry and becomes more asymmetric. This happens as a result of the large change in absorption at higher wavelengths. With an increase in the nonlinear absorption, the valley of the transmittance is severely suppressed and the peak is greatly enhanced, as shown in  Figs.~\textcolor{blue}{\ref{fig:Supp_fig4}(b)} and \textcolor{blue}{(c)}. The larger distortions in the Z-scan signal in the shorter wavelength region is due to the fluctuations in the OPG output in the lower wavelength limit.
\section{6. Nonlinear phase-shift}\label{Section_E}
\begin{figure}[htb]
\centering
\includegraphics[width=6.5cm, height=5.2cm]{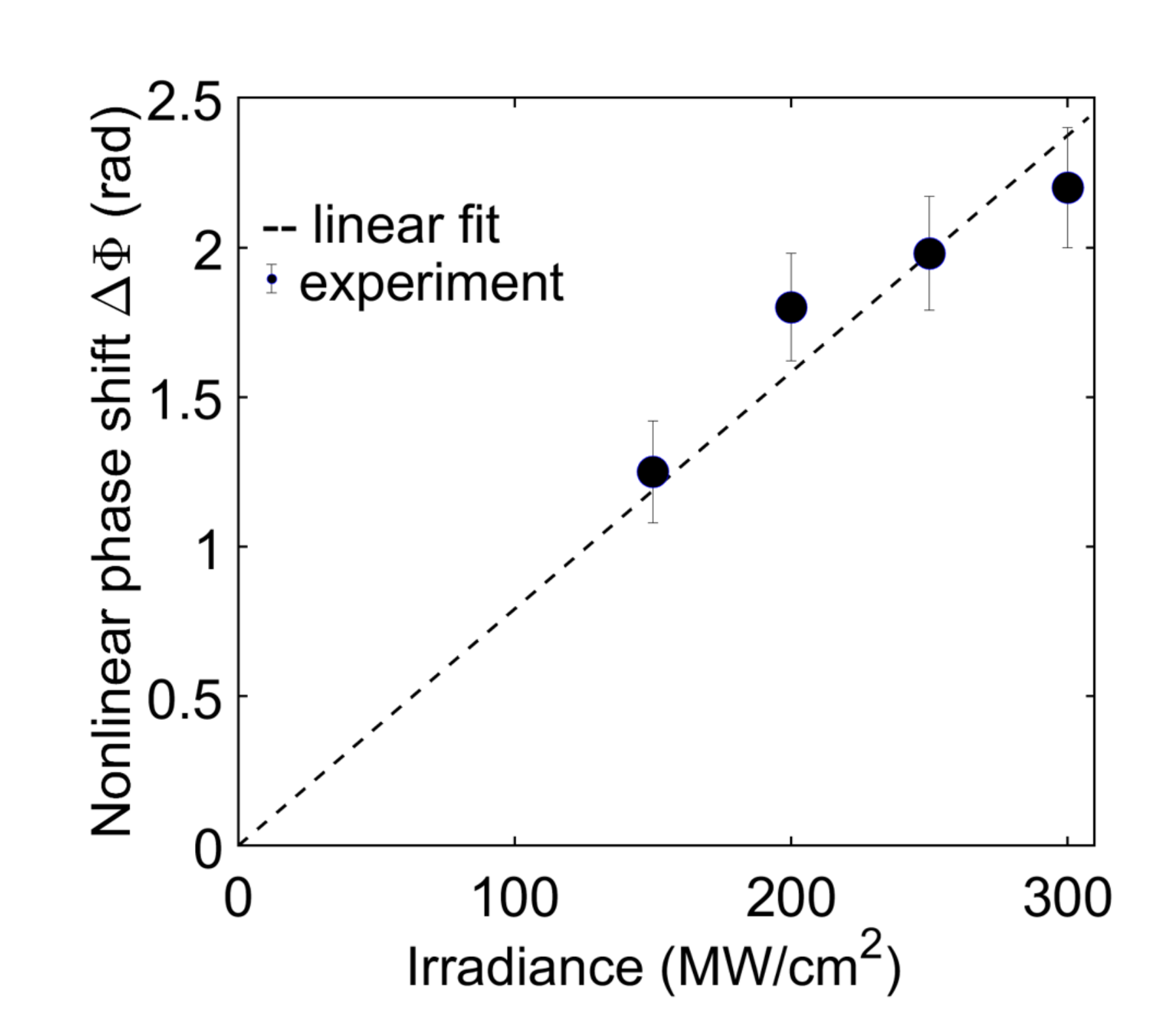}
\caption{(a)~Nonlinear phase shift as a function of the peak irradiance from the Z-scan data at $\lambda$=500~nm. The dashed line is a linear fit to the experimental data.}
\label{fig:supp_fig5}
\end{figure}

The main source of uncertainty in the value of ${n_2}$ is the absolute measurement of the irradiance. A plot of nonlinear phase shift $\Delta\Phi$ versus peak laser irradiance $I_0$ as measured from various Z-scan measurements at the same wavelength is shown in Fig.~\textcolor{blue}{\ref{fig:supp_fig5}}. The nonlinear phase-shift increases linearly with the irradiance on the sample. This in turn can also be used to find the damage threshold of the sample, which is at about 300~$\rm{MW/cm^2}$.

From Fig.~\textcolor{blue}{\ref{fig:supp_fig5}}, we calculated the slope of the linear fit. The nonlinear refractive index $n_2$ is determined using the equation slope= $kn_2L_{\textrm{eff}}$. The value of $n_2$ is therefore found to be ~7.07$~\times10^{-13}~\rm{m^2/W}$ at $\lambda$=500~nm, is very close to the value obtained by fits to individual Z-scan measurement traces (n2= (8.5~$\pm~0.7)~\times10^{-13}$~m$^2$/W). The experimental data is represented by the circles with each point has an error bar. The error bars were evaluated by calculating the standard error over 3 measurements.
\par
A deviation from the linear behaviour could be the result of either higher-order nonlinearities or laser damage to the sample. In this experiment, we determined that the sample undergoes laser damage if the intensity goes above 300~$\rm{MW/cm^2}$. So for all the experimental measurements, we set the intensity of the laser beam below 300~$\rm{MW/cm^2}$ to avoid laser-induced damage.

\end{document}